\documentstyle[preprint,aps]{revtex} 
\begin{document}
\draft
\title{Luttinger Liquid Instability in the One Dimensional {\em t-J} Model }
\author{C. Stephen Hellberg and E. J. Mele}
\address{
Department of Physics, University of Pennsylvania, Philadelphia, PA 19104-6396}
\date{February 25, 1993}
\maketitle

\begin{abstract}
We study the {\em t-J} model in one dimension by numerically projecting the
true ground state from a Luttinger liquid trial wave function.
We find the model exhibits Luttinger liquid behavior for most of the
phase diagram in which interaction strength and density are varied.
However at small densities and high interaction strengths a new
phase with a gap to spin excitations and enhanced superconducting correlations
is found.
We show this phase is a Luther-Emery liquid and study its
correlation functions.
\end{abstract}

The {\em t-J} model was proposed to describe the dynamics of holes
doped into a Mott insulating state
\cite{anderson87,zhang88,gros87}.
Even in one dimension, determining
the complete phase diagram for this apparently simple model
has proven to be quite formidable,
and the ground state structure turns out to be far richer than
initially suspected.
In this paper we combine a variational approach with an exact
ground state projection method to study the properties of this model.

The Hamiltonian for the {\em t-J} model in one dimension can be
written in the subspace of no doubly occupied sites as
\begin{equation}
   H =  -t\sum_{i \sigma}
   ( c_{i\sigma}^{\dagger}c_{i+1\sigma} +  c_{i+1\sigma}^{\dagger}c_{i\sigma})
        +J\sum_{i}
    ( {\bf S}_{i} \cdot {\bf S}_{i+1} - \frac{1}{4} n_i n_{i+1}).\label{tJ:ham}
\end{equation}
The model has been solved exactly only for $J \rightarrow 0$, where
it is equivalent to the $U \rightarrow \infty$ Hubbard model,
and $J = 2t$ \cite{ogata90,kawakami90}.
In both cases
the ground state at arbitrary density belongs to a broad class of interacting
Fermi systems known as Luttinger liquids, which
exhibit power law decay of correlation functions with exponents
characterized by a single parameter
\cite{solyom79,haldane81,emery79}.
Additionally,
for very large $J/t$ the attractive Heisenberg interaction term in
(\ref{tJ:ham}) dominates the
kinetic energy and the model phase separates.

To obtain the rest of phase diagram of the {\em t-J} model,
several numerical approaches have been used.
For example, Ogata, et. al. \cite{ogata91}
have exactly diagonalized this Hamiltonian on a 16 site ring and find
the model behaves as a Luttinger liquid for all values
of $J/t$ below a critical value at which phase separation occurs.
They hypothesized that a third phase of bound singlet pairs
may separate the other phases at very low density but were unable
to resolve this phase with such small system sizes.

In this paper we employ a Luttinger liquid variational wave function to
approximate the ground state of the one dimensional {\em t-J} model
\cite{hellberg91,hellberg91.2,hellberg91.3,hellberg92}
and then use a numerical projection technique to extract
the true ground state from this trial state.
With these methods, we can accurately investigate much larger systems
than attainable by previous techniques.
We confirm that the {\em t-J} model has a Luttinger liquid ground state for
most
of its uniform density phase diagram, and in this region the ground
state is well described by the trial state.
At small densities, however, we find a third phase separating the
Luttinger liquid and phase separated states.
This phase behaves as a Luther-Emery liquid, exhibiting a gap to spin
excitations and enhanced superconducting correlations.

In previous work
\cite{hellberg91.3}
we studied the ground state with a
Luttinger liquid trial state written in the subspace of no doubly occupied
sites
as a Jastrow Slater wave function
\begin{equation}
  \Psi_\nu = \prod_{i<j} |d_{i j}|^{\nu}
  \;\;  S(r^{\uparrow}_{i}) S(r^{\downarrow}_{i}) \label{wf}
\end{equation}
where $S(r_{i}) = Det[e^{ik_{j}r_{i}}]$ is a Slater determinant of single
particle plane wave states
and $d_{ij} = \sin ( \pi (r_i - r_j)/L )$ for a system of size $L$.
The Jastrow factor $\prod_{i<j} |d_{i j}|^{\nu}$ in (\ref{wf}) modulates
the wave function by the distance between all pairs of particles raised
to the power $\nu$, taken as a single variational parameter.
Positive values of $\nu$ induce a smooth correlation hole between all
particles,
while negative values provide an attractive correlation competing with
the Pauli repulsion.
For $\nu < -1/2$ this attraction overcomes the statistical
repulsion, and phase separation occurs.
The long range nature of this Jastrow factor generates the Luttinger
liquid behavior of the wave function
\cite{hellberg92,kawakami92}.
This wave function has been considered in two dimensions where
it also exhibits an algebraic singularity at the Fermi surface
\cite{valenti92,anderson89}.

Applying (\ref{wf}) to the {\em t-J} model one finds
the optimum value of the
variational parameter $\nu$ varies continuously with interaction
strength and density over most of the
phase diagram prior to the critical $J/t$ for phase separation.
However, at small densities we found a third region separating the Luttinger
liquid and phase separated states where the optimized parameter
is pinned at the critical state $\nu = -1/2$.
At this point the many body system in the trial subspace
has infinite compressibility,
which physically cannot extend for a range of interaction strengths.
One concludes that the true ground state here likely lies far from
our variational subspace.
We would like a systematic way of both checking the accuracy of the
trial state where we think it is doing well and determining the
exact ground state in this third region.

In this work, we start with the optimized trial state (\ref{wf})
and project it onto the exact ground state numerically
\cite{liang90,liang90.2,lee93,lee93.2}.
A series of increasingly accurate approximants to the ground state
is generated by
$ |p \rangle = (H - W)^p | \Psi_\nu \rangle$
were $H$ is the Hamiltonian and $W$ is a numerical constant.
These states approach the true ground state for large $p$ provided
$|E_0 - W| > |E_i - W |$ for all excited states $E_i$.
For the {\em t-J} model with $J > 0$ we may choose $W=0$.
In principle this method can be used to project
any trial state not orthogonal to the ground state,
but good initial states give faster convergence and smaller
statistical errors.

To evaluate ground state
expectation values of an arbitrary operator, we calculate
\begin{equation}
\langle O_p \rangle = \frac{\langle p | O | p \rangle}{\langle p |p \rangle}
= \frac{\langle \Psi_\nu | H^pOH^p | \Psi_\nu \rangle}
{\langle \Psi_\nu | H^{2p} | \Psi_\nu \rangle}
\label{Op}
\end{equation}
and take the large $p$ limit.
For sufficiently large powers, the scaling of (\ref{Op}) is dominated
by the contribution from the first excited state overlapping the trial
state.
Thus we can write
\begin{equation}
\langle E_p \rangle = E_0 + \delta \exp (-2\Delta p) + \cdots
\label{energy}
\end{equation}
with $\exp (-\Delta) = |E_1 - W |/|E_0 - W|$.
An operator not commuting with the Hamiltonian has
an additional cross term:
\begin{equation}
\langle O_p \rangle = O_0 + \delta_1 \exp (-\Delta p) +
          \delta_2 \exp (-2\Delta p) + \cdots
\label{operator}
\end{equation}
We use the convergence of the energy
(\ref{energy}) to fix $\Delta$,
and then use (\ref{operator}) to
determine the ground state values of the rest of the observables.

Traditionally (\ref{Op})
has been calculated using a hybrid of two numerical techniques.
First the trial wave function $\Psi$ is sampled with
Variational Monte Carlo to give an ensemble of configurations
$|\alpha\rangle$ \cite{gros87,gros89,ceperley77}.
Then for each $|\alpha\rangle$
the product $H^p$ is sampled
stochastically using a method similar to the Neumann Ulam matrix method
\cite{negele}.
The products are sandwiched to evaluate $\langle H^p O H^p \rangle$
or the normalization  $\langle H^{2p}\rangle$.
This approach throws away much information,
specifically the details of the intermediate states in the
evaluation of each $H^p$.

We developed a much more efficient algorithm for evaluating (\ref{Op})
by combining the two operations.
In usual Variational Monte Carlo a new configuration $|\beta\rangle$
is chosen from a previous configuration $|\alpha\rangle$ with
probability
\begin{equation}
P^{VMC}_{\alpha\rightarrow\beta} =
\min \left( 1, \left| \frac{\Psi_\beta}{\Psi_\alpha} \right|^2 \right)
\label{pvmc}
\end{equation}
After many transitions, this leads to a distribution
of configurations proportional to $|\Psi_\alpha|^2$.
If new configurations are instead chosen with the probability
\begin{equation}
P_{\alpha\rightarrow\beta} =
\frac{1}{z_\alpha} \frac{\Psi_\beta}{\Psi_\alpha} H_{\beta\alpha}
\label{prob}
\end{equation}
with
\begin{equation}
z_\alpha = \sum_{\beta'} \frac{\Psi_{\beta'}}{\Psi_\alpha} H_{\beta'\alpha}
\end{equation}
the distribution for a configuration $|\alpha\rangle$ approaches
$\left| \Psi^2_\alpha / z_\alpha \right|$.
This method of generating new configurations is the same used to evaluate
the products $H^p$, so the operations can be combined.

The algorithm improves on the traditional approach in two ways.
When evaluating a diagonal expectation value, such as
$\langle n(r)n(0) \rangle$ or $\langle S_z(r) S_z(0) \rangle$,
our method evaluates a new $\langle \Psi_\beta | H^pOH^p |\Psi_\alpha\rangle$
at every step of the random walk, so calculations of different
powers $p$ require the same amount of time.
Additionally, for any expectation value, an arbitrary number of different
values of $p$ may be calculated in parallel.
The only disadvantage of our approach is that ergodicity is
violated as $J \rightarrow 0$, and the old method must be used in this limit.
Both methods offer an improvement to Greens Function Monte Carlo
in that exact correlation functions can be calculated
\cite{raedt}.
Since statistical errors grow with increasing $p$, we generally
chose the maximum power to be 10 times the system size.

The phase diagram of the {\em t-J} model determined by our projection
technique is shown in Fig.\ \ref{phase_diagram}.
We see that three distinct phase occur.
For small $J/t$, the ground state is a Luttinger liquid with spin
correlations dominating the long range behavior.
Increasing $J$ suppresses these correlations, and the ground state
passes through the Fermi liquid point of the Luttinger liquid spectrum
at the dashed line.
Above this line the Luttinger liquid has dominant singlet pairing correlations,
and for very large $J/t$ the ground state is phase separated
\cite{ogata91,hellberg91.3,assaad90,putikka92}.
As will be reported in detail elsewhere, in the
Luttinger liquid regime the trial state (\ref{wf}) approximates the exact
ground state very well.

In this work we see clear evidence of a new Luther-Emery liquid
phase (labeled ``Spin Gap'') separating the Luttinger liquid and phase
separated states at small densities \cite{luther74,emery79}.
Unlike all Luttinger states,
this new phase exhibits short range spin correlations,
and thus a gap to spin excitations, while both charge and
singlet pair correlations decay algebraically.
Physically one can view the Luther-Emery liquid as a
translationally invariant
coherent quantum fluid of bound singlet pairs.
The pairs are correlated and can be treated at
a simple level as an interacting fluid of hard core bosons.

Luther-Emery states have been observed in diluted spin models
that exhibit gaps in the saturated state, such as the
{\em t-J} model with Ising anisotropy \cite{pruschke91}
or the next nearest neighbor {\em t-J} model \cite{ogata91.2}.
Additionally this phase is present in the {\em t-J-V} model at quarter filling
\cite{troyer93}.
This work provides the clearest evidence to date
of the spontaneous formation of a Luther-Emery state by doping a gapless
parent state.

A sample spin correlation function in the Luther-Emery phase is
plotted in Fig.\ (\ref{sk}) with the correlation function obtained
from the unprojected trial state shown for comparison.
The variational function exhibits the linear behavior at small wave vectors
characteristic of Luttinger liquids,
while the exact function is quadratic at small $k$ and analytic at all
wave vectors, consistent with exponentially decaying spatial correlations.
We calculate the boundary between the Luttinger and Luther-Emery
states by the crossover from linear to quadratic behavior
at small wave vectors.

More definitive evidence of Luther-Emery behavior can be seen in the
superconducting correlation function plotted in Fig.\ \ref{bk}.
The exponents of the correlations functions in both
Luttinger and Luther-Emery liquids that decay with power laws
can be characterized by
a single parameter $K_\rho \geq 0$ \cite{schulz90,ren91}.
The non-oscillatory part of singlet pair correlation
function decays as
\begin{equation}
\langle b^\dagger (r) b(0) \rangle \propto r^{-\lambda}
\end{equation}
where $b(r) = \frac{1}{\sqrt{2}}
(c_{r\uparrow} c_{r+1\downarrow} - c_{r\downarrow} c_{r+1\uparrow})$.
For Luttinger liquids $\lambda_L = 1 + K_\rho^{-1}$ while
Luther-Emery liquids have
$\lambda_{L-E} = K_\rho^{-1}$.

In Fig.\ \ref{bk}, $b(k)$ diverges logarithmically with system size
as $k \rightarrow 0$ in our trial wave function, which
represents the strongest
divergence possible in a Luttinger liquid state.
However, the true ground
state in this region apparently exhibits a much stronger cusp \cite{lee93}.

Using a finite size scaling analysis of
the divergence in $b(k\rightarrow 0)$
in the projected state,
we can determine the value of this exponent $\lambda$
\cite{assaad90}.
A plot of $\lambda$ showing the transition from Luttinger
to Luther-Emery liquid behavior at density $\delta = 1/6$
is shown in Fig.\ \ref{scaling}.
In the Luttinger regime, $\lambda$ is bounded from below by $1$,
but this bound is clearly violated as the Luther Emery state
is entered.
A continuous variation of $\lambda$ with $J$ as found in this
data would imply a discontinuous jump in $K_\rho$.

It is interesting to note that non-interacting hard core
bosons have $\lambda = 1/2$,
so our singlet pairs have residual repulsive interactions
for $J \lesssim 2.65$ in the Luther-Emery state, while at higher
$J$ the hard core nature of the pairs competes with an effective
attractive interaction \cite{kawakami91}.
The attraction from the Heisenberg term in (\ref{tJ:ham})
in this regime is strong enough to bind singlet pairs but still
insufficient to cause macroscopic phase separation.

Chen and Lee proposed a variational state for this
region by Gutzwiller projecting a sea of non-interacting
bound singlet pairs \cite{lee93.2}.
Their wave function corresponds to a $K_\rho = \infty$ Luther-Emery
state, the critical point of the verge of phase separation which
exhibits a macroscopic superfluid density.
Their calculations of the boundaries of the spin gapped regime
agree remarkably well with ours except at the boundary to phase
separation, which they find occurs at higher $J/t$.
One may speculate that
a potentially more accurate trial state could be generated by
correlating the pairs with a Jastrow factor similar to (\ref{wf}).
This state would exhibit generalized Luther-Emery behavior with
arbitrary $K_\rho$.

In summary, we have investigated the ground state properties of
the {\em t-J} model in one dimension using a numerical technique to
project the exact ground state from a variational Luttinger Liquid
trial state.
We find the model has a surprisingly rich phase diagram.
At lower interaction strengths the variational wave function accurately
describes the Luttinger liquid phase, and at very large $J/t$ the model
phase separates.
However, one finds these phases are separated at low density by a
Luther-Emery quantum dimer liquid phase with short range spin correlations
and enhanced superconducting correlations.

We are grateful to T. K. Lee for valuable discussions.
This work was supported by National Science Foundation Grant
No. DMR 90-08256.

\begin{figure}
\caption{
The phase diagram of the {\em t-J} model as determined in this paper.
The Luther-Emery state is the region labeled ``Spin Gap''.
The dashed line indicates $K_\rho = 1$, the Fermi liquid phase.
Below this line the Luttinger liquid has dominant antiferromagnetic
correlations, while above this line singlet pair correlations decay
with the smallest exponent.
The phase separation boundary is determined by the divergence
of $n(k\rightarrow 0)$, the lower Luther-Emery boundary by the
behavior of $S(k\rightarrow 0)$, and the Fermi liquid line by
$S(k\rightarrow 2k_F)$.
All systems contained at least 100 sites and 10 electrons and holes,
so phase boundaries cannot extent to the extreme densities.
The dotted lines are extrapolations.
}
\label{phase_diagram}
\end{figure}

\begin{figure}
\caption{
The spin-spin correlation function for $J/t = 2.8$ and density
$n = \frac{1}{6}$.
The optimized variational wave function has linear behavior at
small wave vectors while the exact spin correlation turns on quadratically
in $k$.
The system contains 20 electrons on a 120 site lattice.
}
\label{sk}
\end{figure}

\begin{figure}
\caption{
The singlet pair correlation function at $J/t = 2.8$ and density
$n = \frac{1}{6}$.
The $k=0$ cusp is greatly enhanced in the exact ground state.
The system contains 10 electrons on 60 sites.
}
\label{bk}
\end{figure}

\begin{figure}
\caption{
The scaling of the exponent of the $k=0$ superconducting cusp with
interaction strength at density $n = \frac{1}{6}$.
The transition from Luttinger to Luther-Emery liquid states occurs
at $J/t \approx 2.3$
and the system phase separates at $J/t \approx 2.9$.
Luttinger liquids require $\lambda \geq 1$, and non-interacting hard
core bosons have $\lambda = \frac{1}{2}$.
}
\label{scaling}
\end{figure}


\begin{references}
\bibitem{anderson87} P.W. Anderson, Science {\bf 235}, 1196 (1987).
\bibitem{zhang88} F.C. Zhang and T.M. Rice, Phys. Rev. B {\bf 37}, 3759
    (1988).
\bibitem{gros87} C. Gros., R. Joynt, and T.M. Rice, Phys. Rev. B
    {\bf 36}, 381 (1987).
\bibitem{ogata90} M. Ogata and H. Shiba, Phys. Rev. B {\bf 41},
    2326 (1990).
\bibitem{kawakami90} N. Kawakami and S.-K. Yang, Phys. Rev. Lett. {\bf 65},
    2309 (1990).
\bibitem{solyom79} J. S\'{o}lyom, Adv. Phys. {\bf 28}, 201 (1979).
\bibitem{haldane81} F.D.M. Haldane, J. Phys. C. {\bf 14}, 2585 (1981);
    Phys. Rev. Lett. {\bf 45}, 1358 (1980).
\bibitem{emery79}  V.J. Emery, ``Highly Conducting one-dimensional solids'',
    edited by J.T. Devreese et. al., (Plenum, New York, 1979).
\bibitem{ogata91} M. Ogata, M.U. Luchini, S. Sorella, and F.F. Assaad,
    Phys. Rev. Lett. {\bf 66}, 2388 (1991).
\bibitem{hellberg91} C.S. Hellberg and E.J. Mele, Phys. Rev. B {\bf 44},
    1360 (1991).
\bibitem{hellberg91.2} C.S. Hellberg and E.J. Mele, Int. J. Mod. Phys. B
    {\bf 5}, 1791 (1991).
\bibitem{hellberg91.3} C.S. Hellberg and E.J. Mele, Phys. Rev. Lett. {\bf 67},
    2080 (1991).
\bibitem{hellberg92} C.S. Hellberg and E.J. Mele, Phys. Rev. Lett. {\bf 68},
    3111 (1992).
\bibitem{kawakami92} N. Kawakami and P. Horsch, Phys. Rev. Lett. {\bf 68},
    3110 (1992).
\bibitem{valenti92} R. Valent\'{\i} and C. Gros, Phys. Rev. Lett. {\bf 68},
    2402 (1992); C. Gros and R. Valent\'{\i}, preprint.
\bibitem{anderson89} P.W. Anderson, B.S. Shastry, and D. Hristopulos,
    Phys. Rev. B {\bf 40}, 8939 (1989).
\bibitem{liang90} S. Liang, Phys. Rev. Lett. {\bf 64}, 1597 (1990).
\bibitem{liang90.2} S. Liang, Phys. Rev. B {\bf 42}, 6555 (1990).
\bibitem{lee93} Y.C. Chen and T.K. Lee, preprint.
\bibitem{lee93.2} Y.C. Chen and T.K. Lee, preprint.
\bibitem{gros89} C. Gros, Ann. Phys. (N.Y.) {\bf 189}, 53 (1989).
\bibitem{ceperley77} D. Ceperley, G.V. Chester, and M.H. Kalos, Phys. Rev. B
    {\bf 16}, 3081 (1977).
\bibitem{negele} J.W. Negele and H. Orland, ``Quantum Many-Particle Systems''
    (Addison-Wesley, New York, 1987).
\bibitem{raedt} H. De Raedt and W. von der Linden, in ``The Monte Carlo Method
    in Condensed Matter Physics'', edited by K. Binder (Springer-Verlag,
Berlin,
    1992).
\bibitem{assaad90} F.F. Assaad and D. W\"{u}rtz,
    Phys. Rev. B {\bf 44}, 2681 (1991).
\bibitem{putikka92} W.O. Putikka, M.U. Luchini, and T.M. Rice,
    Phys. Rev. Lett. {\bf 68}, 538 (1992).
\bibitem{luther74} A. Luther and V.J. Emery, Phys. Rev. Lett., {\bf 33}
    589 (1974).
\bibitem{pruschke91} T. Pruschke and H. Shiba, Phys. Rev. B
    {\bf 44}, 205 (1991).
\bibitem{ogata91.2} M. Ogata, M.U. Luchini, and T.M. Rice, Phys. Rev. B
    {\bf 44}, 12083 (1991).
\bibitem{troyer93} M. Troyer, H. Tsunetsugu, T.M. Rice, J. Riera,
    and E. Dagotto, preprint.
\bibitem{schulz90} H.J. Schulz, Phys. Rev. Lett. {\bf 64}, 2831 (1990).
\bibitem{ren91}  Y. Ren and P.W. Anderson, preprint 1990.
\bibitem{kawakami91} N. Kawakami and S.-K. Yang, Phys. Rev. Lett. {\bf 67},
    2493 (1991).
\end{references}
\end{document}